\documentclass{natureprintstyle2}
\usepackage{graphicx,amsmath}
\usepackage{astjnlabbrev-nature} 

\usepackage{amssymb}

\title{Star clusters: Anything but simple}

\author{Richard de Grijs$^{1,2,3}$}
\begin{document}

\maketitle

\begin{affiliations}
 \item Kavli Institute for Astronomy and Astrophysics, Peking
   University, Yi He Yuan Lu 5, Hai Dian District, Beijing 100871,
   China; e-mail: grijs@pku.edu.cn\\
 \item Department of Astronomy, Peking University, Yi He Yuan Lu 5,
   Hai Dian District, Beijing 100871, China\\
 \item Discipline Scientist, International Space Science
   Institute--Beijing\\
\end{affiliations}

\begin{abstract} 
The heated debate on the importance of stellar rotation and age
spreads in massive star clusters has just become hotter by throwing
stellar variability into the mix.
\end{abstract}

A quiet revolution has been sweeping the field of star-cluster
astrophysics. A decade ago, we were reasonably convinced that we
understood the formation and evolution of the massive, well-populated
star clusters that can be used as a statistical tool for studies of
stellar evolution. Groups of stars characterized by a common age and
chemical composition were considered `simple stellar populations',
given that all of their stars had presumably formed from the same
progenitor molecular gas cloud at approximately the same
time. Admittedly, the oldest galactic building blocks, the globular
clusters, were known to exhibit evidence of multiple stellar
generations\cite{ref1}, but clusters younger than a few billion years
appeared to obey our simple models. Fast forward a decade, and we now
know that the majority of 1--3 billion-year-old star clusters in the
nearest galaxies, the Magellanic Clouds, are anything but
simple. Indeed, writing in {\it The Astrophysical Journal Letters},
Ricardo Salinas and co-workers show that a significant population of
pulsating stars can have a measurable effect on our interpretation of
stellar evolution within such clusters\cite{ref2}.

Deviations from the simple stellar population model show up most
readily in a cluster's colour--magnitude diagram. This type of plot is
the observational counterpart to the theoretical Hertzsprung--Russell
diagram, which relates the temperatures (or colours) of the cluster's
stars to their luminosities. Instead of being randomly distributed,
the stars tend to lie on bands. Most stars, including the Sun, belong
to the `main sequence', when they are fusing hydrogen into helium in
their cores. By mapping a stellar population in this manner, it is
possible to estimate the age of the stars in a given
cluster.

Most of the `intermediate-age' clusters in the Magellanic Clouds
exhibit extended regions in colour-magnitude space\cite{ref3,ref4} at
the `main-sequence turn-off'---the evolutionary phase where stars have
exhausted their core hydrogen---but still on the `main sequence',
before commencing hydrogen fusion in a thin shell surrounding their
cores. Single-aged, single-metallicity stellar populations would,
instead, exhibit narrow ridgelines and sharp turn-offs. Initial
explanations for the extended main-sequence turn-off areas suggested
that massive clusters might have continued forming stars for some time
following a cluster's initial burst of star formation\cite{ref5}. This
would also generate a range of metal abundances over time as new
generations of stars formed from the chemically enhanced debris of
their progenitors. This idea has lost traction in recent years with
the realization that star clusters may be composed of coeval stellar
populations after all, but whose stars might be characterized by a
range of rotation rates\cite{ref4}.

In the classical `instability strip' in the Hertzsprung--Russell
diagram, stars become unstable and exhibit pulsations because of
cyclical abundance changes of singly and doubly ionized helium in
their atmospheres\cite{ref6} (Fig. 1). It crosses the main sequence
for A- and F-type stars, that is, for stars with masses ranging from
approximately 1.5 to 2.5 solar masses. Conventional stellar evolution
theory implies that such stars occupy the main-sequence turn-offs in
coeval star clusters with ages of about 1--3 billion years. The
majority of main-sequence turn-off stars are stable, even those
located inside the instability strip. Yet, certain stellar types
exhibit photometric variability, including the rapidly oscillating
peculiar A-type (`roAp') stars, SX Phoenicis and $\delta$ Scuti
variables. The $\delta$ Scuti variables show periodic luminosity
changes ranging from 30 minutes to 8 hours, which are driven by both
radial and non-radial (wave-like) pulsations on the stellar surface.

\begin{figure}
\includegraphics[width=\columnwidth]{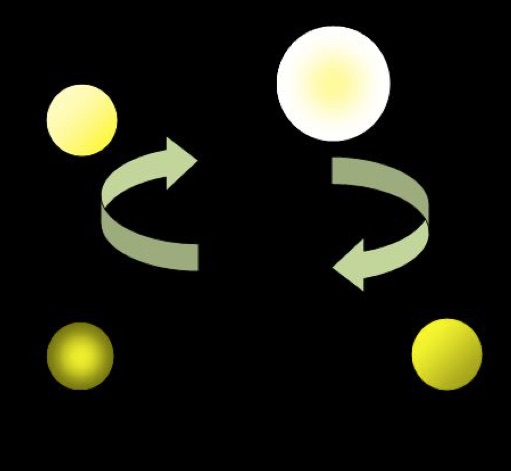}
\begin{center}
\caption{Heart of brightness. Variable stars such as the $\delta$
  Scuti variables change their luminosity and temperature in a
  periodic fashion, thus appearing to pulsate. In the dimmest phase,
  the outer shell is rich in He$^{2+}$ and is opaque, so radiation
  from within gets trapped. As it warms, the star expands and
  cools. The He$^{2+}$ then converts to He$^+$, which is more
  transparent, allowing the heat to escape. As the star continues to
  cool, the expansion stops, and eventually reverses under the star's
  own gravity. ({\it Figure adapted from Antonine Education})}
\end{center}
\label{F1}
\end{figure}

Salinas {\it et al.}\cite{ref2} point out that the effects of the
luminosity and colour changes of $\delta$ Scuti stars in the
main-sequence turn-off area have been completely ignored. The authors
analyse theoretical colour--magnitude diagrams, varying both the
fraction of the main-sequence stars residing in the instability strip
which are actually pulsating variables---a ratio known as the
`incidence'---and their maximum photometric amplitudes. Their first
important conclusion states that the density of cluster stars near the
observational ridgeline (or, alternatively, the theoretical isochrone)
decreases as the incidence increases from 10\% to 50\%, with the
distribution becoming as much as 50\% wider for the highest incidence.

Second, and perhaps most interesting, their analysis implies that the
extent of the main-sequence turn-off region owing to the presence of
$\delta$ Scuti stars is maximal for cluster ages around 2 billion
years. Clusters younger than 1 billion years or those older than 2.5
billion years are not affected because of the complex interplay
between the location of $\delta$ Scuti stars on the main sequence and
its age-dependent overlap with the instability strip. This fresh
insight is eerily similar to the results from a recent independent
analysis which considered the apparent internal cluster age spread
implied by the extent of the main-sequence turn-off as a function of
cluster age, reaching a maximum at an age of 1.5--1.7 billion
years\cite{ref7}.

The results of Salinas and co-workers are intriguing and offer
significant food for thought. They naturally explain the observed
absence both of broadened subgiant branches in the colour--magnitude
diagrams\cite{ref8} and of extended red clumps\cite{ref9}. Yet, the
actual incidence of $\delta$ Scuti variables in single-aged star
clusters is unknown, so that current estimates are necessarily based
on the properties of their counterparts among the Milky Way's field
stellar population---perhaps not the best comparison
sample. Observational data to confirm or reject these novel 
\vfill\eject\noindent
insights
are, unfortunately, challenging to obtain. As there are no suitable
young or intermediate-age clusters available in our Milky Way, we
would need to secure time-series observations at high spatial
resolution of 1--3 billion-year-old star clusters in the Magellanic
Clouds. This approach would require {\sl Hubble Space Telescope}
capabilities; even with their adaptive optics capability turned on,
the European Southern Observatory's Very Large Telescope cannot attain
the resolution needed, given the Magellanic Clouds' location deep in
the southern hemisphere and the correspondingly large air column
affecting such observations. Therefore, the viability of the Salinas
{\it et al.} proposal remains to be tested, but at least the field can
now move forward again.

\end{document}